\documentclass[prd,reprint,showpacs,showkeys]{revtex4-1}
\usepackage{amsfonts}
\usepackage{amssymb}
\usepackage{amsmath}
\usepackage{graphicx}
\usepackage[font={footnotesize,it}]{caption}

\begin{document}

\title{A topological metric in 2+1-dimensions}
\author{S. Habib Mazharimousavi}
\email{habib.mazhari@emu.edu.tr}
\author{M. Halilsoy}
\email{mustafa.halilsoy@emu.edu.tr}
\affiliation{Department of Physics, Eastern Mediterranean University, Gazima\~{g}usa,
Turkey. }
\date{\today }

\begin{abstract}
Real-valued triplet of scalar fields as source gives rise to a metric which
tilts the scalar, not the light cone, in 2+1-dimensions. The topological
metric is static, regular and characterized by an integer $\kappa =\pm 1,\pm
2,...$. The problem is formulated as a harmonic map of Riemannian manifolds
in which the integer $\kappa $ equals to the degree of the map.
\end{abstract}

\maketitle

\section{Introduction}

The topic of metrical kinks has a long history in general relativity \cite{1}
which declined recently toward oblivion. On the other hand in a broader
sense interest in topological aspects in non-linear field theory, for a
number of reasons, remains ever alive. Although these emerge mostly in flat $%
3+1-$dimensional spacetime with the advent of higher / lower dimensions the
same topological concepts may find applications in these cases as well.

The aim of this paper is to revisit this subject in $2+1-$dimensions.
Motivation for this lies in part by the discovery of a cosmological black
hole \cite{2} which became a center of attraction in this particular
dimension. Does a topological metric also make a black hole? The answer to
this question turns negative, at least in our present study. The derived
metric is sourced by a triplet of scalar fields $\phi ^{a}\left( r,\theta
\right) ,$ ($a=1,2,3,$ is the internal index) which satisfies the
constraint\ $\left( \phi ^{a}\right) ^{2}=1,$ with topological properties.
Let us add that besides the topological solution our system of triplet
fields admits special solutions including black holes. Our interest,
however, will be focused on the topological one. Unlike the geometrical kink
metrics \cite{1} which tilt the light cone, leading to closed timelike
curves, our topological metric tilts the scalar field along its range. The
metric admits an integer, $\kappa =\pm 1,\pm 2,...$ which can be interpreted
(following Ref. \cite{3}) as the topological charge / homotopy class. The
total energy is a multiple of $\left\vert \kappa \right\vert $ and relation
with the harmonic maps of Riemannian manifolds \cite{4} suggests that $%
\kappa $ is at the same time the degree of the map. Let us note that Ref. 
\cite{3} was extended shortly afterwards in a detailed analysis by Clement 
\cite{5}. In search for a topological particle interpretation Clement gave
models and exact solutions in lower dimensions that can further be
generalized to higher dimensions. That is, in $n+1-$dimensions, $n-$scalar
fields model can be considered in a flat spacetime without much difficulty.
Similar considerations in a curved spacetime with gravitation naturally adds
its own complications. In particular, the solution given in \cite{5} for $%
2+1-$dimensions with gravitation relates closely to our study, which happens
yet to be different from what has been considered here.

\section{The Formalism}

In the $2+1-$dimensional spacetime 
\begin{equation}
ds^{2}=-A\left( r\right) dt^{2}+\frac{dr^{2}}{B\left( r\right) }%
+r^{2}d\theta ^{2}
\end{equation}%
we choose the action ($16\pi G=c=1$)%
\begin{equation}
I=\int d^{3}x\sqrt{-g}\left( R-\frac{1}{2}\left( \nabla \phi ^{a}\right)
^{2}-\frac{\lambda }{2}\left( \left( \phi ^{a}\right) ^{2}-1\right) \right)
\end{equation}%
in which 
\begin{equation}
\phi ^{a}\left( r,\theta \right) =\left( 
\begin{array}{c}
\sin \alpha \left( r\right) \cos \beta \left( \theta \right) \\ 
\sin \alpha \left( r\right) \sin \beta \left( \theta \right) \\ 
\cos \alpha \left( r\right)%
\end{array}%
\right) .
\end{equation}%
Our notation stands as follows: $R$ is the Ricci scalar, $A\left( r\right) $%
, $B\left( r\right) $ and $\alpha \left( r\right) $ are functions of $r,$ $%
\beta \left( \theta \right) $ is a function of $\theta $ and $\lambda \left(
r,\theta \right) $ is a Lagrange multiplier. $\phi ^{a}\left( r,\theta
\right) $ transforms under the symmetry group $O\left( 3\right) $ and
satisfies 
\begin{equation}
\phi ^{a}\phi ^{a}=1.
\end{equation}%
Variational principle yields the field equation%
\begin{equation}
\square \phi ^{a}=\lambda \phi ^{a}
\end{equation}%
in which $\square $ stands for the covariant Laplacian, and the constraint
condition (4).

In the sequel we shall make the choice 
\begin{equation}
\beta \left( \theta \right) =\kappa \theta
\end{equation}%
with $\kappa =\pm $(integer) for uniqueness condition. This reduces the
action effectively, modulo the time sector, to 
\begin{equation}
I=\int rdr\sqrt{\frac{A}{B}}\left( R-\frac{1}{2}B\alpha ^{\prime 2}-\frac{%
\kappa ^{2}}{2r^{2}}\sin ^{2}\alpha \right)
\end{equation}%
in which a prime stands for $\frac{d}{dr}.$ With the energy-momentum tensor 
\begin{equation}
T_{\mu }^{\nu }=\frac{1}{2}\left[ \left( \partial _{\mu }\phi ^{a}\right)
\left( \partial ^{\nu }\phi ^{a}\right) -\frac{1}{2}\left( \nabla \phi
^{a}\right) ^{2}\delta _{\mu }^{\nu }\right]
\end{equation}%
variation with respect to $\alpha \left( r\right) $ and Einstein equations 
\begin{equation}
G_{\mu }^{\nu }=T_{\mu }^{\nu }
\end{equation}%
we obtain the following equations%
\begin{equation}
\left( r\sqrt{AB}\alpha ^{\prime }\right) ^{\prime }=\frac{\kappa ^{2}}{2r}%
\sqrt{\frac{A}{B}}\sin 2\alpha
\end{equation}%
\begin{equation}
\frac{-2B^{\prime }}{r}=B\alpha ^{\prime 2}+\frac{\kappa ^{2}}{r^{2}}\sin
^{2}\alpha
\end{equation}%
\begin{equation}
\frac{2BA^{\prime }}{rA}=B\alpha ^{\prime 2}-\frac{\kappa }{r^{2}}\sin
^{2}\alpha
\end{equation}%
\begin{equation}
2A^{\prime \prime }-\frac{A^{\prime 2}}{A}+\frac{A^{\prime }B^{\prime }}{B}=-%
\frac{A}{B}\left( B\alpha ^{\prime 2}-\frac{\kappa ^{2}}{r^{2}}\sin
^{2}\alpha \right) .
\end{equation}%
This system of differential equations admits a number of particular
solutions. As examples we give the followings.

\subsection{A black hole solution}

This is obtained by 
\begin{equation}
\alpha =\frac{\pi }{2}
\end{equation}%
\begin{equation}
A\left( r\right) =B\left( r\right) =C_{0}-\frac{\kappa ^{2}}{2}\ln r
\end{equation}%
where $C_{0}$ is an integration constant that can be interpreted as mass.
The scalar field triplet takes the form%
\begin{equation}
\phi ^{a}\left( \theta \right) =\left( 
\begin{array}{c}
\cos \kappa \theta \\ 
\sin \kappa \theta \\ 
0%
\end{array}%
\right)
\end{equation}%
which is effectively a doublet of scalars. Ricci scalar of this solution
reads%
\begin{equation}
R=\frac{\kappa ^{2}}{2r^{2}}
\end{equation}%
which is singular at $r=0.$ Event horizon $r_{h}$ of the resulting black
hole is 
\begin{equation}
r_{h}=e^{2C_{0}/\kappa ^{2}}
\end{equation}%
so that it is characterized by the index $\kappa .$ Similarly the Hawking
temperature also is stamped by the integer $\kappa ^{2}$. Clearly this is a
different situation from the Ba\~{n}ados-Teitelboim-Zanelli (BTZ) black hole 
\cite{2}, where the parameter, i.e. cosmological constant (and electric
charge) are not integers.

\subsection{The topological solution}

Our system of equations (10-13) admits a solution with the choice $A\left(
r\right) =1.$ Accordingly, Eq. (10) reduces to the Sine-Gordon equation%
\begin{equation}
\left( 2\alpha \right) _{uu}=\sin \left( 2\alpha \right)
\end{equation}%
where%
\begin{equation}
e^{2u}=\left( \frac{r_{0}}{r}\right) ^{2}-1
\end{equation}%
in which $r_{0}$ is a constant that will be set simply to $r_{0}=1.$ In the
new variable 
\begin{equation}
\rho =\tanh ^{-1}r.
\end{equation}%
the solution for $\alpha \left( \rho \right) $ and $B\left( \rho \right) $
become%
\begin{equation}
\alpha \left( \rho \right) =2\tan ^{-1}\left( \frac{1}{\sinh \rho }\right)
\end{equation}%
\begin{equation}
B\left( \rho \right) =\frac{\kappa ^{2}}{\cosh ^{4}\rho }
\end{equation}%
so that the resulting $2+1-$dimensional line element takes the form%
\begin{equation}
ds^{2}=-dt^{2}+\frac{d\rho ^{2}}{\kappa ^{2}}+\tanh ^{2}\rho d\theta ^{2},
\end{equation}%
which can be cast through the transformation 
\begin{equation}
\sinh \rho =\left( \frac{r}{r_{0}}\right) ^{\kappa }
\end{equation}%
into the form%
\begin{equation}
ds^{2}=-dt^{2}+\frac{\left( dr^{2}+r^{2}d\theta ^{2}\right) }{r^{2}\left(
1+\left( \frac{r_{0}}{r}\right) ^{2\kappa }\right) }
\end{equation}%
in which $r_{0}$ is a constant parameter. The solution given in \cite{5} for
the particular analytic function $\psi \left( z\right) =z=re^{i\theta }$
(see Eq. (4.13) of \cite{5}) can be expressed by 
\begin{equation}
ds^{2}=-dt^{2}+\frac{\left( dr^{2}+r^{2}d\theta ^{2}\right) }{\left( 1+\frac{%
r^{2}}{2\nu ^{2}}\right) ^{2\chi \nu ^{2}}}
\end{equation}%
in which $\chi $ and $\nu $ are constant parameters. It can be easily
checked that for $\kappa =1,$ (26) coincides, after a time scaling, with
(27) upon choice of the parameters 
\begin{equation}
\chi =\frac{1}{r_{0}^{2}}=\frac{1}{2\nu ^{2}}.
\end{equation}%
For $\kappa \neq 1,$ the two metrics are different. Our metric (26) has also
no correspondence with the point particle solution of \cite{5} (see the
Appendix of \cite{5}). The analogy is valid only for the extended model of
particles and confined only to $\kappa =1.$

Our metric (24) represents a regular, non-black hole, static spacetime. The
non-zero geometrical quantities are%
\begin{equation}
\text{Ricci scalar: }R=\frac{4\kappa ^{2}}{\cosh ^{2}\rho }
\end{equation}%
\begin{equation}
\text{Kretschmann scalar: }K=\frac{1}{2}R^{2}
\end{equation}%
and%
\begin{equation}
R_{\mu \nu }R^{\mu \nu }=K
\end{equation}%
with the energy-momentum tensor 
\begin{equation}
T_{\mu }^{\nu }=-\frac{R}{2}\delta _{\mu }^{0}\delta _{0}^{\nu }.
\end{equation}%
As a result our triplet of scalar fields take the form%
\begin{equation}
\phi ^{a}\left( \rho ,\theta \right) =\left( 
\begin{array}{c}
\left( 
\begin{array}{c}
\cos \kappa \theta  \\ 
\sin \kappa \theta 
\end{array}%
\right) \frac{2\sinh \rho }{\cosh ^{2}\rho } \\ 
\frac{\sinh ^{2}\rho -1}{\cosh ^{2}\rho }%
\end{array}%
\right) .
\end{equation}%
This leads for $\rho =0$, to%
\begin{equation}
\phi ^{a}\left( 0,\theta \right) =\left( 
\begin{array}{c}
0 \\ 
0 \\ 
-1%
\end{array}%
\right) 
\end{equation}%
while for $\rho \rightarrow \infty $ we have 
\begin{equation}
\phi ^{a}\left( \infty ,\theta \right) =\left( 
\begin{array}{c}
0 \\ 
0 \\ 
1%
\end{array}%
\right) .
\end{equation}%
It is observed that between $0\leq \rho <\infty $ the angle $\alpha \left(
\rho \right) $ shifts from $-1$ to $+1,$ which amounts to the case of
one-kink. It should also be remarked that 'kink' herein is used in the sense
of flip of the $\phi ^{3}$ component of the triplet, not in the sense of
light cone tilt. The energy density of the kink is maximum at $\rho =0,$
which decays asymptotically whose energy $E_{\kappa }$ is 
\begin{equation}
E_{\kappa }=\int_{0}^{\infty }\int_{0}^{2\pi }\left( -T_{t}^{t}\right) \sqrt{%
-g}d\rho d\theta =4\pi \left\vert \kappa \right\vert .
\end{equation}

\subsection{The harmonic map formulation}

We wish to add, for completeness that the $\alpha \left( r\right) $ equation
can be described as a harmonic map, between two Riemannian manifolds $M$ and 
$M^{\prime }$ \cite{4}%
\begin{equation}
f^{A}:M\rightarrow M^{\prime }
\end{equation}%
which are defined by%
\begin{eqnarray}
M^{\prime } &:&ds^{\prime 2}=d\alpha ^{2}+\sin ^{2}\alpha d\beta ^{2} \\
&=&g_{AB}^{\prime }df^{A}df^{B},\text{ (}A,B=1,2\text{)}  \notag
\end{eqnarray}%
and%
\begin{eqnarray}
M &:&ds^{2}=\frac{d\rho ^{2}}{\kappa ^{2}}+\tanh ^{2}\rho d\theta ^{2} \\
&=&g_{ab}dx^{a}dx^{b},\text{ (}a,b=1,2\text{).}  \notag
\end{eqnarray}%
The energy functional of the map is defined by 
\begin{equation}
E\left( f^{A}\right) =\int g_{AB}^{\prime }\frac{df^{A}}{dx^{a}}\frac{df^{B}%
}{dx^{b}}g^{ab}\sqrt{g}d^{2}x.
\end{equation}%
which yields, upon variation the equation for $\alpha \left( \rho \right) $.
Note that in this map we consider a priori that $\alpha =\alpha \left( \rho
\right) $ and $\beta =\beta \left( \theta \right) $. The degree of harmonic
map $\left( d\right) $ is defined in an orthonormal frame $\left\{
x^{i}\right\} $ by 
\begin{equation}
d=\frac{1}{2\pi }\int d^{2}x\sin \alpha \left\vert \frac{\partial \left(
\alpha ,\beta \right) }{\partial \left( x_{1},x_{2}\right) }\right\vert
=\kappa
\end{equation}%
which equals to the topological charge \cite{3}.

Although the maps in the original work of Eells and Sampson \cite{4} were
considered between unit spheres (in particular $S^{2}\rightarrow S^{2}$) in
the present problem our map is from $R^{2}\rightarrow S^{2}.$ We must add
that the method of harmonic maps was proposed long ago as a model for a
non-linear field theory \cite{6}. Einstein's equations of general relativity
also followed from a harmonic map formulation \cite{7}. The isometries of
the M' metric serves to generate new solutions from known solutions \cite{8}%
. Unfortunately the non-compact and singular manifolds of general relativity
create serious handicaps which prevented a wider application of the concept
of degrees of the maps once they are formulated in harmonic forms.

\section{Conclusion}

In conclusion we comment that topological properties of field theory were
well-defined in a flat space background. Due to the singular and non-compact
manifolds of general relativity these concepts found no simple applications
in a curved spacetime. In this note, similar to the contribution of Ref. 
\cite{5} we have shown that at least in the $2+1-$dimensional spacetime the
problem can be overcome. The source of our metric is provided by a triplet
of scalar fields which may find applications as multiplets of scalar fields
in higher-dimensions. It has been shown that the triplet source gives rise
to other solutions, such as black holes, besides the topological metric. The
interesting feature of such a black hole is that event horizon and as a
result the Hawking temperature due to the integer $\kappa $ takes discrete
multiple values of certain value. As a final remark let us add that it
should be interesting to extend our model to $3+1-$dimensions with
multi-scalar fields. The technical problems such as the non-linear
superposition of Sine-Gordon solutions in a curved space leading to the
'multi-kink' metric remains to be seen.

\textbf{Acknowledgement:} We wish to thank the anonymous referee for
directing our attention to the Ref. \cite{5}.

\bigskip


\begin{thebibliography}{99}
\bibitem{1} D. Finkelstein and C. W. Misner, Annals of Physics, \textbf{6},
230 (1959);

J. G. Williams and R. K. P. Zia, J. Phys. A: Math. Nucl. Gen. \textbf{6}, 1
(1973);

J. G. Williams, J. Phys. A: Math. Nucl. Gen. \textbf{7}, 1871 (1974);

D. Finkelstein and G. McCollum, Jour. Math. Phys. \textbf{16}, 2250 (1975);

M. Vasilic and T. Vukasinac, Class. Quantum Grav. \textbf{12}, 1995 (1996);

J. G. Williams, Gen. Rel. Grav. \textbf{23}, 1995 (1996);

T. Kl\"{o}sch and T. Strobl, Phys. Rev. D \textbf{57}, 1034 (1998);

J. G. Williams, Gen. Rel. Grav. \textbf{30}, 27 (1998).

\bibitem{2} M. Ba\~{n}ados, C. Teitelboim, J. Zanelli, Phys. Rev. Lett. 
\textbf{69}, 1849 (1992).

M. Ba\~{n}ados, M. Henneaux, C. Teitelboim and J. Zanelli, Phys. Rev. D 
\textbf{48}, 1506 (1993).

\bibitem{3} J. Honerkamp, A. Patani, M. Schlindwein and Q. Shafi, Lett. al
Nuovo Cimento, \textbf{15}, 97 (1975).

\bibitem{4} J. Eells and J. H. Sampson, Amer. J. Math. \textbf{86}, 109
(1964).

\bibitem{5} G. Clement, Nucl. Phys. B \textbf{l14, }437 (1976).

\bibitem{6} C. W. Misner, Phys. Rev. D \textbf{18}, 4510 (1978).

\bibitem{7} Y. Nutku, Ann. Inst. Henri Poincare, XXI, 175 (1974).

\bibitem{8} M. Halilsoy, Phys. Lett. A \textbf{84}, 404 (1981).
\end{thebibliography}
\end{document}